\documentstyle[12pt,epsf,epsfig]{article}
\newcount\timecount
\newcount\hours \newcount\minutes  \newcount\temp \newcount\pmhours

\hours = \time
\divide\hours by 60
\temp = \hours
\multiply\temp by 60
\minutes = \time
\advance\minutes by -\temp
\def\hour{\the\hours}
\def\minute{\ifnum\minutes<10 0\the\minutes
            \else\the\minutes\fi}
\def\clock{
\ifnum\hours=0 12:\minute\ AM
\else\ifnum\hours<12 \hour:\minute\ AM
      \else\ifnum\hours=12 12:\minute\ PM
            \else\ifnum\hours>12
                 \pmhours=\hours
                 \advance\pmhours by -12
                 \the\pmhours:\minute\ PM
                 \fi
            \fi
      \fi
\fi
}

\def\monthname{\relax\ifcase\month 0/\or January\or February\or
   March\or April\or May\or June\or July\or August\or September\or
   October\or November\or December\else\number\month/\fi}

\def\bold#1{\setbox0=\hbox{$#1$}%
     \kern-.025em\copy0\kern-\wd0
     \kern.05em\copy0\kern-\wd0
     \kern-.025em\raise.0433em\box0 }

\def\ga{\mathrel{\raise.3ex\hbox{$>$\kern-.75em\lower1ex\hbox{$\sim$}}}}
\def\la{\mathrel{\raise.3ex\hbox{$<$\kern-.75em\lower1ex\hbox{$\sim$}}}}
\def\gev{{\rm \, Ge\kern-0.125em V}}
\def\tev{{\rm \, Te\kern-0.125em V}}
\def\beq{\begin{equation}}
\def\eeq{\end{equation}}

\def\m12{m_{1\!/2}}

\def\Yi{\eta^{\ast}_{11} \left( \frac{y_{i}}{2} g' Z_{\chi 1} + 
        g T_{3i} Z_{\chi 2} \right) + \eta^{\ast}_{12} 
        \frac{g m_{q_{i}} Z_{\chi 5-i}}{2 m_{W} B_{i}}}

\def\Xii{\eta^{\ast}_{11} 
        \frac{g m_{q_{i}}Z_{\chi 5-i}^{\ast}}{2 m_{W} B_{i}} - 
        \eta_{12}^{\ast} e_{i} g' Z_{\chi 1}^{\ast}}

\def\Wi{\eta_{21}^{\ast}
        \frac{g m_{q_{i}}Z_{\chi 5-i}^{\ast}}{2 m_{W} B_{i}} -
        \eta_{22}^{\ast} e_{i} g' Z_{\chi 1}^{\ast}}

\def\Vi{\eta_{22}^{\ast} \frac{g m_{q_{i}} Z_{\chi 5-i}}{2 m_{W} B_{i}}
        + \eta_{21}^{\ast}\left( \frac{y_{i}}{2} g' Z_{\chi 1}
        + g T_{3i} Z_{\chi 2} \right)}

\def\zthree{\delta_{1i} [g Z_{\chi 2} - g' Z_{\chi 1}]}

\def\zfour{\delta_{2i} [g Z_{\chi 2} - g' Z_{\chi 1}]}

\begin{document}
\begin{titlepage}
\pagestyle{empty}
\baselineskip=21pt
\rightline{hep-ph/0001005}
\rightline{CERN--TH/2K-001}
\rightline{UMN--TH--1834/2K}
\rightline{TPI--MINN--2K/01}
\vskip 0.2in
\begin{center}
{\large{\bf Re-Evaluation of the Elastic Scattering of Supersymmetric
Dark Matter}}
\end{center}
\begin{center}
\vskip 0.2in
{{\bf John Ellis}$^1$, {\bf Andrew Ferstl}$^2$ and {\bf Keith
A.~Olive}$^{2,3}$}\\
\vskip 0.1in
{\it
$^1${TH Division, CERN, Geneva, Switzerland}\\
$^2${School of Physics and Astronomy,
University of Minnesota, Minneapolis, MN 55455, USA}\\
$^3${Theoretical Physics Institute,
University of Minnesota, Minneapolis, MN 55455, USA}}\\
\vskip 0.2in
{\bf Abstract}
\end{center}
\baselineskip=18pt \noindent
We examine the cross sections for the elastic scattering of
neutralinos $\chi$ on nucleons $p,n$, as functions of
$m_\chi$ in the constrained minimal supersymmetric standard model.
We find narrow bands of possible values of the cross section, that are
considerably lower than some previous estimates.  The constrained model
is based on the minimal supergravity-inspired framework for the MSSM, with
universal scalar and gaugino masses $m_0, m_{1/2}$, and 
$\mu$ and the MSSM Higgs masses treated as dependent parameters. We
explore
systematically the region of the $(m_{1/2}, m_0)$ plane where LEP and
other accelerator constraints are respected, and the relic neutralino
density lies in the range $0.1 \le \Omega_{\chi} h^2 \le 0.3$ preferred by
cosmology. We update previous discussions of both the spin-independent and
-dependent scattering matrix elements on protons and neutrons, using
recent analyses of low-energy hadron experiments. 

\vfill
\leftline{CERN--TH/2K-001}
\leftline{January 2000}
\end{titlepage}
\baselineskip=18pt
\section{Introduction}

One of the most promising candidates for the cold dark matter
believed to pervade the Universe is the lightest supersymmetric
particle (LSP)~\cite{EHNOS}, commonly expected to be the lightest
neutralino $\chi$,
which is stable in the minimal supersymmetric extension of the Standard
Model (MSSM) with conserved $R$ parity~\cite{MSSM}. The quantum stability
of the gauge
hierarchy suggests that sparticles weigh less than about
1~TeV~\cite{hierarchy}, which is also the range favoured for a cold
dark matter particle~\cite{Dim}, and there
are indeed generic domains of the MSSM parameter space in which the
relic LSP density falls within the range $0.1 \le \Omega_\chi h^2 \le 0.3$
favoured by astrophysics and cosmology~\cite{EHNOS}. The unsuccessful
laboratory
searches for sparticles impose non-trivial constraints on the MSSM
parameter space, suggesting that the LSP $\chi$ is mainly a $U(1)$
gaugino (Bino)~\cite{EFGOS}.

Many non-accelerator strategies to search for cosmological relic
neutralinos have been proposed~\cite{GJK}, including indirect searches for
products of their annihilations in free space or inside
astrophysical bodies, and direct searches for their scattering on
target nuclei in low-background underground laboratories~\cite{GW}. The
rates
for such experiments typically have larger uncertainties than those
for producing sparticles at accelerators, since they involve some
astrophysical and/or cosmological uncertainties as well as those
due to simulations of the signatures, over and above the common
uncertainties in the MSSM parameters. Nevertheless, such dark
matter searches offer interesting prospects for beating
accelerators to the discovery of supersymmetry, particularly during
the coming years before the LHC enters operation.

In this paper, we embark on a programme to clarify the extents
of the uncertainties in searches for supersymmetric dark matter, by
re-evaluating the rates to be expected for the elastic scattering
of relic LSPs on protons and neutrons~\cite{EF,etal}. Large ranges for
these rates
are often quoted~\cite{bigguys}, reflecting general explorations of the
MSSM parameter
space. Our approach is to establish as accurately as possible a
baseline set of predictions based on the most plausible
assumptions commonly used in constrained MSSM phenomenology, such as
universality in the soft supersymmetry-breaking parameters as
suggested by minimal supergravity models, and requiring the
cosmological relic density to lie within the range favoured by
astrophysics and cosmology, namely $0.1 \le \Omega_\chi h^2 \le 0.3$.
These assumptions can and should be questioned, but they are well
motivated and good candidates for default options in analyses of
the MSSM and cold dark matter.

In the course of this re-evaluation of elastic $\chi - p,n$ scattering
cross sections, we re-analyze the relevant spin-independent and
spin-dependent matrix elements of scalar densities and axial currents
in protons and neutrons. We update previous analyses using further
information from chiral symmetry~\cite{leut,Cheng}, low-energy $\pi -
p,n$
scattering~\cite{Gasser} and deep-inelastic lepton-nucleon
scattering~\cite{Mallot}. We include a
discussion of uncertainties in the values of the scalar and
axial-current matrix elements.

We perform a systematic scan of the region of the $m_0, m_{1/2}$ parameter
space of the MSSM with supergravity-inspired universality that is
consistent with accelerator constraints and yields a cosmological
relic density within the favoured range $0.1 \le \Omega_\chi \le
0.3$~\cite{EFOSi}.
We treat $\mu$ as a dependent parameter (modulo a sign ambiguity), and our
results are not very
sensitive to $A$. We order our results in terms of $m_\chi$ which closely
tracks $m_{1/2}$. For any given choice of $m_\chi, \tan\beta$ and the sign
of
$\mu$, we find a relatively narrow band of possible
cross sections, reflecting the fact that the accelerator and
cosmological constraints~\cite{EFGOS,EFOSi} favour a predominant $U(1)$
gaugino (Bino)
composition for the LSP. Our results fall considerably below many
of the possible predictions in the literature~\cite{bigguys}, and may
discourage
some faint-hearted experimentalists. However, we think they
provide a realistic estimate of the target sensitivity required
for an experiment to have a good chance of success.

\section{Theoretical Framework}

We review in this Section the theoretical framework we use in the context
of the MSSM~\cite{MSSM}. The neutralino LSP is the
lowest-mass eigenstate combination of the Bino ${\tilde B}$, Wino $\tilde W$
and Higgsinos ${\tilde H}_{1,2}$, whose mass matrix $N$ is
diagonalized by a matrix $Z$: $diag(m_{\chi_1,..,4}) = Z^* N Z^{-1}$.
The composition of the lightest neutralino may be written as
\begin{equation}
\chi = Z_{\chi 1}\tilde{B} + Z_{\chi 2}\tilde{W} +
Z_{\chi 3}\tilde{H_{1}} + Z_{\chi 4}\tilde{H_{2}}
\label{id}
\end{equation}
As already mentioned, we assume universality at the
supersymmetric GUT scale for the 
$U(1)$ and $SU(2)$ gaugino masses: $M_{1,2} = m_{1/2}$, so that 
$M_1 = \frac{5}{3}\tan^2{\theta_{W}}M_{2}$ at the electroweak scale.
We denote by $\tan{\beta}$ the ratio of Higgs vacuum expectation
values, and $\mu$ is the Higgsino mass-mixing parameter. We also assume
GUT-scale universality for the soft supersymmetry-breaking scalar masses
$m_0$, for the Higgs bosons as well as the squarks and sleptons. We further
assume GUT-scale universality for the soft supersymmetry-breaking
trilinear terms $A$.
Our treatment of the sfermion mass matrices $M$ follows~\cite{FFO1}.
As discussed there, the sfermion mass-squared matrix is diagonalized by a
matrix
$\eta$: $diag(m^2_1, m^2_2) \equiv
\eta M^2 \eta^{-1}$, which can be parameterized 
for each flavour $f$ by an angle $ \theta_{f} $ and phase $\gamma_f$:
\begin{equation}
\left( \begin{array}{cc}
\cos{\theta_{f}} & \sin{\theta_{f}} e^{i \gamma_{f}} \nonumber \\
-\sin{\theta_{f}} e^{-i \gamma_{f}} & \cos{\theta_{f}}
\end{array} \right) 
\hspace{0.5cm}
 \equiv 
\hspace{0.5cm}
 \left( \begin{array}{cc}
\eta_{11} & \eta_{12} \nonumber \\
\eta_{21} & \eta_{22}
\end{array} \right)
\label{defineeta}
\end{equation}
As a simplification, we neglect CP violation in this paper, so that
$\gamma_f = 0$ and there are no CP-violating phases in the neutralino
mass matrix, either. We treat $m_{1/2}, m_0, A$ and
$\tan\beta$ as free parameters, and $\mu$ and the pseudoscalar Higgs mass
$m_A$ as dependent parameters specified by the electroweak vacuum
conditions, which we calculate using $m_t = 175$~GeV~\footnote{We
have checked that varying $m_t$ by $\pm 5$~GeV has a negligible
effect on our results.}.

The MSSM Lagrangian leads to the following low-energy effective four-fermi
Lagrangian suitable for describing elastic $\chi$-nucleon
scattering~\cite{FFO1}:
\begin{equation}
{\cal L} = \bar{\chi} \gamma^\mu \gamma^5 \chi \bar{q_{i}} 
\gamma_{\mu} (\alpha_{1i} + \alpha_{2i} \gamma^{5}) q_{i} +
\alpha_{3i} \bar{\chi} \chi \bar{q_{i}} q_{i} + 
\alpha_{4i} \bar{\chi} \gamma^{5} \chi \bar{q_{i}} \gamma^{5} q_{i}+
\alpha_{5i} \bar{\chi} \chi \bar{q_{i}} \gamma^{5} q_{i} +
\alpha_{6i} \bar{\chi} \gamma^{5} \chi \bar{q_{i}} q_{i}
\label{lagr}
\end{equation}
This Lagrangian is to be summed over the quark generations, and the 
subscript $i$ labels up-type quarks ($i=1$) and down-type quarks
($i=2$).  The terms with coefficients $\alpha_{1i}, \alpha_{4i},
\alpha_{5i}$ and $\alpha_{6i}$ make contributions to the 
elastic scattering cross section that are velocity-dependent,
and may be neglected for our purposes. In fact, if the CP
violating phases are absent as assumed here, $\alpha_5 = \alpha_6 =
0$~\cite{FFO2}. The coefficients relevant for our discussion are: 

\begin{eqnarray}
\alpha_{2i} & = & \frac{1}{4(m^{2}_{1i} - m^{2}_{\chi})} \left[
\left| Y_{i} \right|^{2} + \left| X_{i} \right|^{2} \right] 
+ \frac{1}{4(m^{2}_{2i} - m^{2}_{\chi})} \left[ 
\left| V_{i} \right|^{2} + \left| W_{i} \right|^{2} \right] \nonumber \\
& & \mbox{} - \frac{g^{2}}{4 m_{Z}^{2} \cos^{2}{\theta_{W}}} \left[
\left| Z_{\chi_{3}} \right|^{2} - \left| Z_{\chi_{4}} \right|^{2}
\right] \frac{T_{3i}}{2}
\label{alpha2}
\end{eqnarray}
and
\begin{eqnarray}
\alpha_{3i} & = & - \frac{1}{2(m^{2}_{1i} - m^{2}_{\chi})} Re \left[
\left( X_{i} \right) \left( Y_{i} \right)^{\ast} \right] 
- \frac{1}{2(m^{2}_{2i} - m^{2}_{\chi})} Re \left[ 
\left( W_{i} \right) \left( V_{i} \right)^{\ast} \right] \nonumber \\
& & \mbox{} - \frac{g m_{qi}}{4 m_{W} B_{i}} \left[ Re \left( 
\zthree \right) D_{i} C_{i} \left( - \frac{1}{m^{2}_{H_{1}}} + 
\frac{1}{m^{2}_{H_{2}}} \right) \right. \nonumber \\
& & \mbox{} +  Re \left. \left( \zfour \right) \left( 
\frac{D_{i}^{2}}{m^{2}_{H_{2}}}+ \frac{C_{i}^{2}}{m^{2}_{H_{1}}} 
\right) \right]
\label{alpha3}
\end{eqnarray}
where
\begin{eqnarray}
X_{i}& \equiv& \Xii \nonumber \\
Y_{i}& \equiv& \Yi \nonumber \\
W_{i}& \equiv& \Wi \nonumber \\
V_{i}& \equiv& \Vi
\label{xywz}
\end{eqnarray}
where $y_i, T_{3i}$ denote hypercharge and isospin, and
\begin{eqnarray}
\delta_{1i} = Z_{\chi 3} (Z_{\chi 4}) &,& \delta_{2i} = Z_{\chi 4}
(-Z_{\chi 3}), \nonumber \\
B_{i} = \sin{\beta} (\cos{\beta}) &,& A_{i} = \cos{\beta} ( -\sin{\beta}), 
\nonumber \\
C_{i} = \sin{\alpha} (\cos{\alpha}) &,& D_{i} = \cos{\alpha} ( -
\sin{\alpha}) 
\label{moredefs}
\end{eqnarray}
for up (down) type quarks. We denote by $m_{H_2} < m_{H_1}$
the two scalar Higgs masses, and $ \alpha $ denotes the Higgs mixing
angle~\footnote{We note that (\ref{alpha3}) is taken from
\cite{FFO2} and corrects an error in~\cite{FFO1},
and that (\ref{alpha2}, \ref{alpha3}) agree with~\cite{GJK,EF} and the
published version of~\cite{CIN}.}.

\section{Hadronic Matrix Elements}

The elastic cross section for scattering off a nucleus can be
decomposed into a scalar (spin-independent) part obtained from the
$\alpha_{2i}$ term in (\ref{lagr}), and a spin-dependent part
obtained from the $\alpha_{3i}$ term. Each of these can be
written in terms of the cross sections for elastic scattering
for scattering off individual nucleons, as we now review
and re-evaluate.

The scalar part of the
cross section can be written as
\begin{equation}
\sigma_{3} = \frac{4 m_{r}^{2}}{\pi} \left[ Z f_{p} + (A-Z) f_{n} 
\right]^{2}
\label{si}
\end{equation}
where $m_r$ is the reduced LSP mass,
\begin{equation}
\frac{f_{p}}{m_{p}} = \sum_{q=u,d,s} f_{Tq}^{(p)} 
\frac{\alpha_{3q}}{m_{q}} +
\frac{2}{27} f_{TG}^{(p)} \sum_{c,b,t} \frac{\alpha_{3q}}{m_q}
\label{f}
\end{equation}
and $f_{n}$ has a similar expression.  The parameters
$f_{Tq}^{(p)}$  are defined
by
\begin{equation}
m_p f_{Tq}^{(p)} \equiv \langle p | m_{q} \bar{q} q | p \rangle
\equiv m_q B_q
\label{defbq}
\eeq
whilst $ f_{TG}^{(p)} = 1 - \sum_{q=u,d,s} f_{Tq}^{(p)} $~\cite{SVZ}.
We observe that only the products $m_q B_q$, the ratios of the quark masses
$m_q$ and the ratios of the scalar matrix elements $B_q$ are
invariant under renormalization and hence physical quantities.

We take the ratios of the quark masses from~\cite{leut}:
\beq
{m_u \over m_d} = 0.553 \pm 0.043 , \qquad
{m_s \over m_d} = 18.9 \pm 0.8
\eeq
In order to determine the ratios of the $B_q$ and the products $m_q B_q$
we use information from chiral symmetry applied to baryons.
Following~\cite{Cheng}, we have:
\beq
z \equiv {B_u - B_s \over B_d - B_s} =
{m_{\Xi^0} + m_{\Xi^-} -m_p -m_n \over 
m_{\Sigma^+} + m_{\Sigma^-} -m_p -m_n}
\label{cheng}
\eeq
Substituting the experimental values of these baryon masses, we find
\beq
z = 1.49
\label{chengvalue}
\eeq
with an experimental error that is negligible compared with others
discussed below. Defining
\beq
y \equiv {2 B_s \over B_d + B_u},
\label{defy}
\eeq
we then have 
\beq
{B_d \over B_u} = {2 + ((z - 1) \times y) \over 2 \times z - ((z - 1) \times
y)}
\label{bdoverbu}
\eeq
The experimental value of the $\pi$-nucleon $\sigma$ term
is~\cite{Gasser}:
\beq
\sigma \equiv {1 \over 2} (m_u + m_d) \times (B_d + B_u) = 45 \pm 8~{\rm MeV}
\label{sigma}
\eeq 
and octet baryon mass differences may be used to estimate
that~\cite{Gasser}
\beq
\sigma = {\sigma_0 \over (1 - y)}: \qquad \sigma_0 = 36 \pm 7~{\rm MeV}
\label{sigma0}
\eeq
Comparing (\ref{sigma}) and (\ref{sigma0}), we find a central value of
$y = 0.2$, to which we assign an error $\pm 0.1$, which yields
\beq
{B_d \over B_u} = 0.73 \pm 0.02
\eeq
The formal error in $y$ derived from (\ref{sigma}) and (\ref{sigma0}) is
actually $\pm$0.2, which would double the error in $B_d/B_u$. We have
chosen the smaller uncertainty because we consider a value
of y in excess of 30\% rather unlikely. However, we do illustrate
later by one example the potential consequences of a
larger error in $y$.

The numerical magnitudes of the
individual renormalization-invariant products
$m_q B_q$ and hence the
$f_{Tq}^{(p)}$ may now be determined:
\beq
f_{Tu}^{(p)} = 0.020 \pm 0.004, \qquad f_{Td}^{(p)} = 0.026 \pm 0.005,
\qquad f_{Ts}^{(p)} = 0.118 \pm 0.062
\label{pinput}
\eeq 
where essentially all the error in $f_{Ts}^{(p)}$ arises from the
uncertainty in $y$. The corresponding values for the neutron are
\beq
f_{Tu}^{(n)} = 0.014 \pm 0.003, \qquad f_{Td}^{(n)} = 0.036 \pm 0.008,
\qquad f_{Ts}^{(n)} = 0.118 \pm 0.062.
\eeq
It is clear already that the difference between the scalar parts
of the cross sections for scattering off protons and neutrons must be
rather small.

The spin-dependent part of the elastic $\chi$-nucleus cross section can be
written as
\begin{equation}
\sigma_{2} = \frac{32}{\pi} G_{F}^{2} m_{r}^{2} \Lambda^{2} J(J + 1)
\label{sd}
\end{equation}
where $m_{r}$ is again the reduced neutralino mass, $J$ is the spin 
of the nucleus, and
\begin{equation}
\Lambda \equiv \frac{1}{J} (a_{p} \langle S_{p} \rangle + a_{n} \langle
S_{n} \rangle)
\label{lamda}
\end{equation}
where
\begin{equation}
a_{p} = \sum_{i} \frac{\alpha_{2i}}{\sqrt{2} G_{f}} \Delta_{i}^{(p)}, 
a_{n} = \sum_{i} \frac{\alpha_{2i}}{\sqrt{2} G_{f}} \Delta_{i}^{(n)}
\label{a}
\end{equation}
The factors $\Delta_{i}^{(p,n)}$ parametrize the quark spin content of the
nucleon. A recent global analysis of QCD sum rules for the $g_1$
structure functions~\cite{Mallot}, including ${\cal O}(\alpha_s^3)$
corrections,
corresponds formally to the values
\beq
\Delta_{u}^{(p)} = 0.78 \pm 0.02, \qquad \Delta_{d}^{(p)} = -0.48 \pm
0.02,
\qquad \Delta_{s}^{(p)} = - 0.15 \pm 0.02
\label{spincontent}
\eeq
whilst perturbative QCD fits to the data for $g_1$ tend to give
broader ranges~\cite{Mallot}. In our numerical analysis, we double the
formal
errors in (\ref{spincontent}) to $\pm 0.04$, essentially 100\% correlated
for the three quark flavours. In the case of the neutron, we have
$\Delta_{u}^{(n)} = \Delta_{d}^{(p)}, 
\Delta_{d}^{(n)} = \Delta_{u}^{(p)}$, and 
$\Delta_{s}^{(n)} = \Delta_{s}^{(p)}$.

\section{Cosmological and Experimental Constraints}

The domain of MSSM parameter space that we explore in this paper is
that defined in~\cite{EFOSi}. Several convergent measures of
cosmological parameters~\cite{triangle} suggest that the cold dark matter
density $\Omega_{CDM} = 0.3 \pm 0.1$ and that the Hubble
expansion rate $H \equiv h \times 100$~km/s/Mpc: $h = 0.7 \pm 0.1$,
leading to the preferred range $0.1 \le \Omega_{CDM} h^2 \le 0.3$.
The upper limit on $\Omega_{CDM}$ can be translated directly into
the corresponding upper limit on $\Omega_\chi$. However, it is
possible that there is more than one component in the cold dark
matter, so that $\Omega_\chi < \Omega_{CDM}$, opening up the
possibility that $\Omega_\chi < 0.1$. 
Although the MSSM parameters which lead to $\Omega_\chi < 0.1$ tend
to give larger elastic scattering cross sections, the detection
rate also must be reduced because of the corresponding reduction
in the density of LSPs in the
Galactic halo. Here we shall neglect this possibility,
assuming instead that essentially all the cold dark matter is composed
of LSPs, so that $\Omega_\chi \ge 0.1$.

For the calculation of the relic LSP density, we follow~\cite{EFOSi},
where coannihilations between $\chi$ and the sleptons $\tilde \ell$,
particularly the lighter stau $\tilde \tau_1$, were shown to play an
important role. As we discuss in more detail later, $m_\chi$
depends essentially on $m_{1/2}$, and coannihilation increases by
a factor $\sim 2$ the cosmological upper limit on $m_{1/2}$ to $\sim 1400$~GeV,
allowing $m_\chi \la 600$~GeV. At this upper limit on $m_{1/2}$, 
there is a unique allowed value of $m_0 \sim 350$~GeV, but for lower
values of $m_{1/2}$ the width of the 
allowed range of $m_0$ expands, reaching $50 \la m_0 \la 150$~GeV when 
$m_{1/2} \sim 200$~GeV. At this value of $m_{1/2}$ and scanning across
the cosmological range in $m_0$, we find
that $m_\chi \sim 80$~GeV, with a small variation by $ \sim 0.4$~GeV.
These numbers are not very sensitive to
$\tan\beta$ in the range from 3 to 10 studied in~\cite{EFOSi}
and here, nor are they very sensitive to the chosen value of $A$. 

The lower limit on $m_{1/2}$ and hence $m_\chi$ depends on the
sparticle search limits provided by LEP~\cite{EFGOS}. The most essential
of these
for our current purposes are those provided by the experimental
lower limits on the lighter chargino mass $m_{\chi^\pm}$ and the
lighter scalar Higgs mass $m_{H_2}$. A lower limit $m_{\chi^\pm} \ge 95$~GeV
was assumed in~\cite{EFOSi}: unsuccessful chargino searches during
higher-energy runs of LEP have now increased this lower limit to
$m_{\chi^\pm} \ge 100$~GeV~\cite{newLEP}, which does not reduce very much
the range allowed in~\cite{EFOSi}. 

The impact of the recently-improved lower
limits on the Higgs mass~\cite{newLEP} is potentially more significant,
particularly
for $\tan\beta = 3$, as displayed in Figs.~6 and 7 of~\cite{EFOSi}. The
present experimental lower limit for $\tan\beta = 3$ is probably
$m_{H_2} > 105$ to 109~GeV~\cite{newLEP}. The $m_{H_2}$ contours shown in
Figs.~6a,b and
7b of~\cite{EFOSi} were not calculated with the most recent two-loop MSSM
code~\cite{newHiggs},
so we take the $m_{H_2} = 100$~GeV lines in~\cite{EFOSi} as indicative
constraints. These correspond to $m_{1/2} \sim 340 (720)$~GeV for $\mu > (<)
0$, corresponding in turn to $m_\chi \ga 140 (310)$~GeV. On the other
hand, for $\tan\beta = 10$, the LEP lower limit on $m_{H_2}$ is
considerably weaker than $100$~GeV~\cite{newLEP}, and hence
does not constrain significantly the allowed parameter space, as seen in
Figs.~6c,d and 7c of~\cite{EFOSi}. 

We note in passing that requiring our present electroweak
vacuum to be stable against transitions to a lower-energy state in which
electromagnetic charge and colour are broken (CCB)~\cite{AF} would divide
the
parameter regions allowed in~\cite{EFOSi} into two parts: one at
large $m_{1/2}$ and the other at small $m_{1/2}$ and relatively large
$m_0$. We do not implement the CCB constraint in our analysis, since it
may be considered optional. Nor do we implement any constraint due to
the observed rate of $b \rightarrow s \gamma$ decay~\cite{bsg}, but it is
well known that
this reduces very substantially the parameter space allowed for $\mu < 0$.

\section{Results}

As discussed above, we scan the cosmologically preferred set of
parameters which yield $0.1 \le \Omega_\chi h^2 \le 0.3$ and are
consistent with the recent LEP accelerator bounds. For each value of 
$\tan \beta$ and sign of $\mu$, we vary $m_{1/2}$ and $m_0$ over all the
allowed range. As default, we
choose $A_0 = -m_{1/2}$ in most of our computations. Then, using the
hadronic inputs described in section 3, we compute separately the
spin-dependent
and scalar contributions from the $\alpha_2$ and $\alpha_3$ 
coefficients, respectively, to
the elastic scattering of LSPs on both protons and neutrons.

In Figure 1, we show the resulting spin-dependent elastic cross section
as a function of the LSP mass, $m_\chi$. Although it
is barely discernible, the thicknesses of the central curves in the
panels show the ranges in the cross section for fixed $m_\chi$
that are induced by varying $m_0$. At large
$m_\chi$ where coannihilations are important, the range in the allowed
values of $m_0$ is small and 
particularly little variation in the cross section is
expected.  The shaded regions in this and the 
following figures show the effects of the
uncertainties in the input values of the $\Delta^{(p)}_i$ 
(\ref{spincontent}).  In Figure 1a, for $\tan \beta = 3, \mu<0$, we see
at small $m_\chi$
the effect of a cancellation induced by the difference in signs between
$\Delta_u$ and $\Delta_{d,s}$.  Cancellations are possible for the other
values of $\tan \beta$ and sign of $\mu$, but not in the preferred range
of $m_{1/2}$ and $m_0$ used here. Aside from the cancellation, the
spin-dependent cross section peaks at about $10^{-4}$ pb and drops 
rapidly as $m_\chi$ increases.

\begin{figure}
\vspace*{-1.8in}
\hspace*{-.45in}
\begin{minipage}{8in}
\epsfig{file=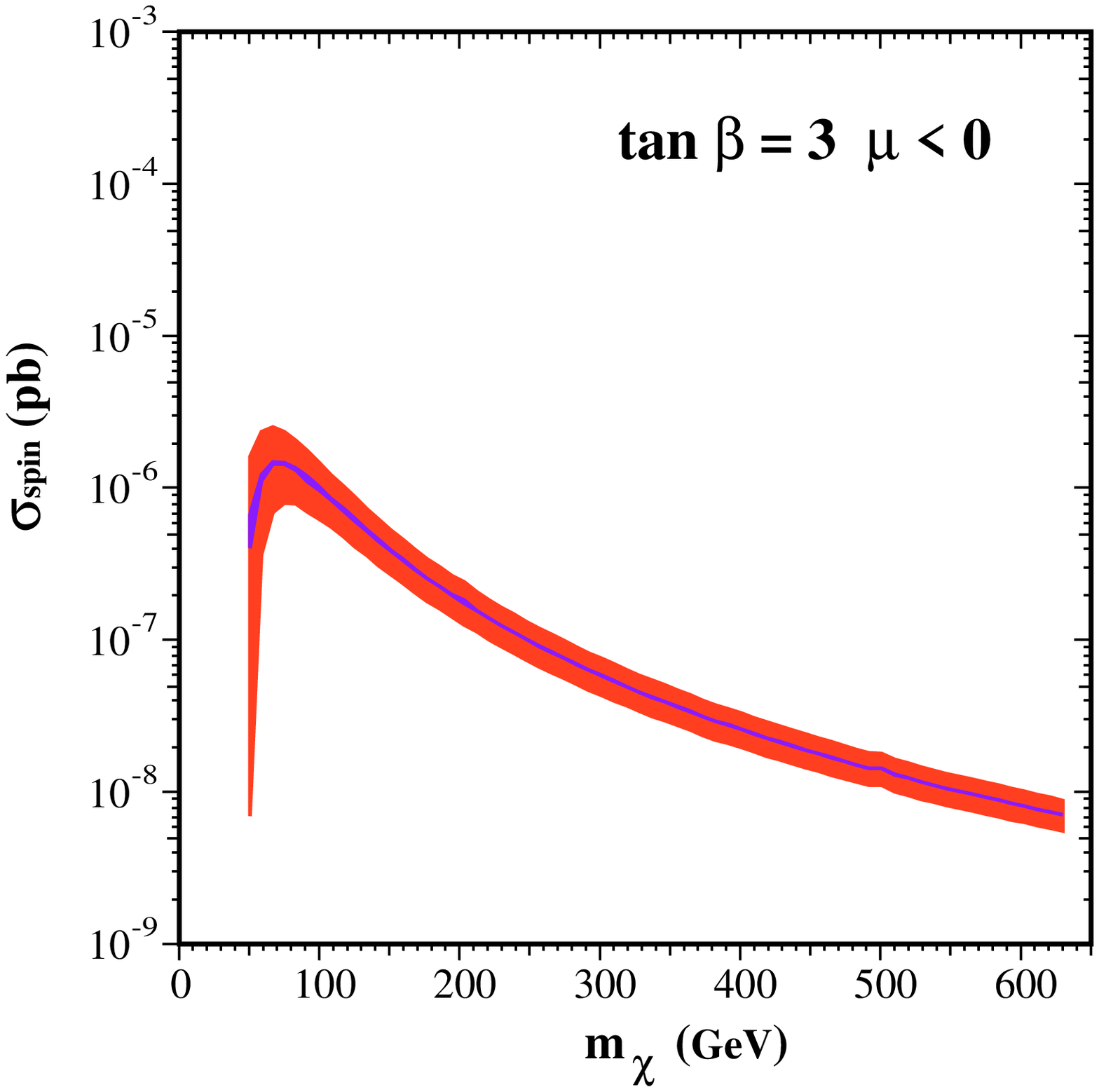,height=5.5in} 
\hspace*{-.40in}\epsfig{file=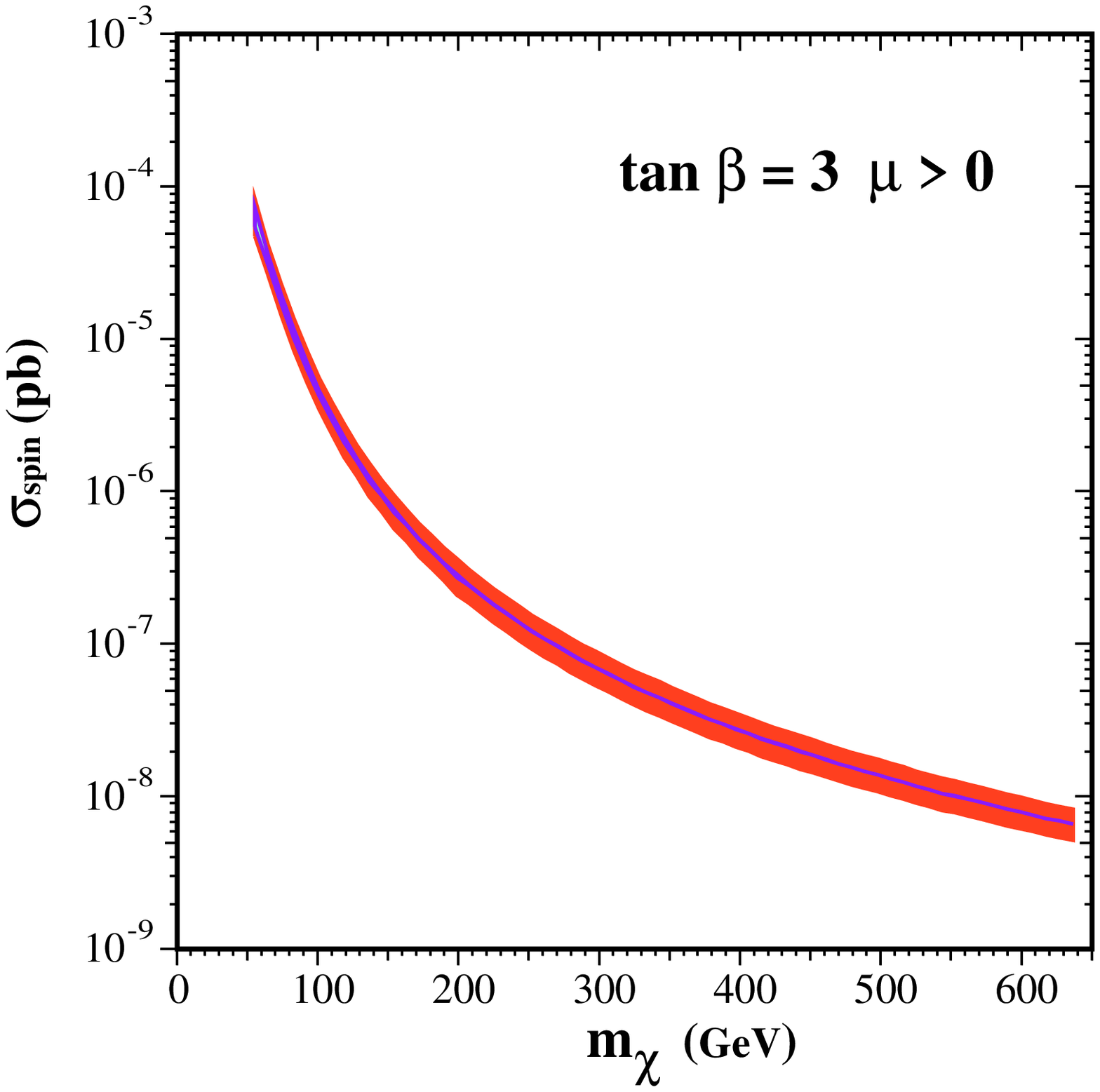,height=5.5in} \hfill
\end{minipage}
\begin{minipage}{8in}
\vspace*{-1.9in}
\hspace*{-.45in}
\epsfig{file=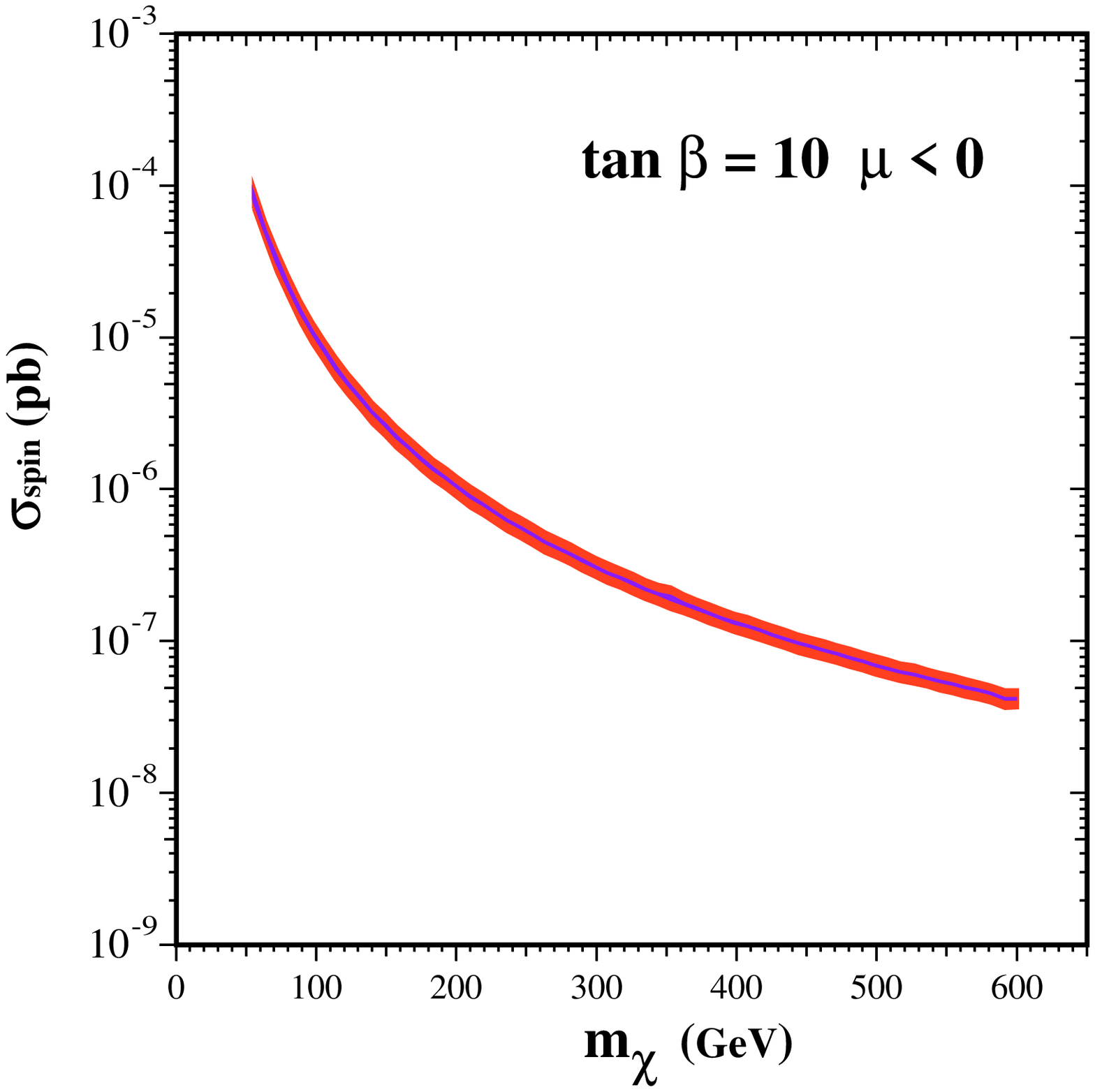,height=5.5in} 
\hspace*{-0.4in}
\epsfig{file=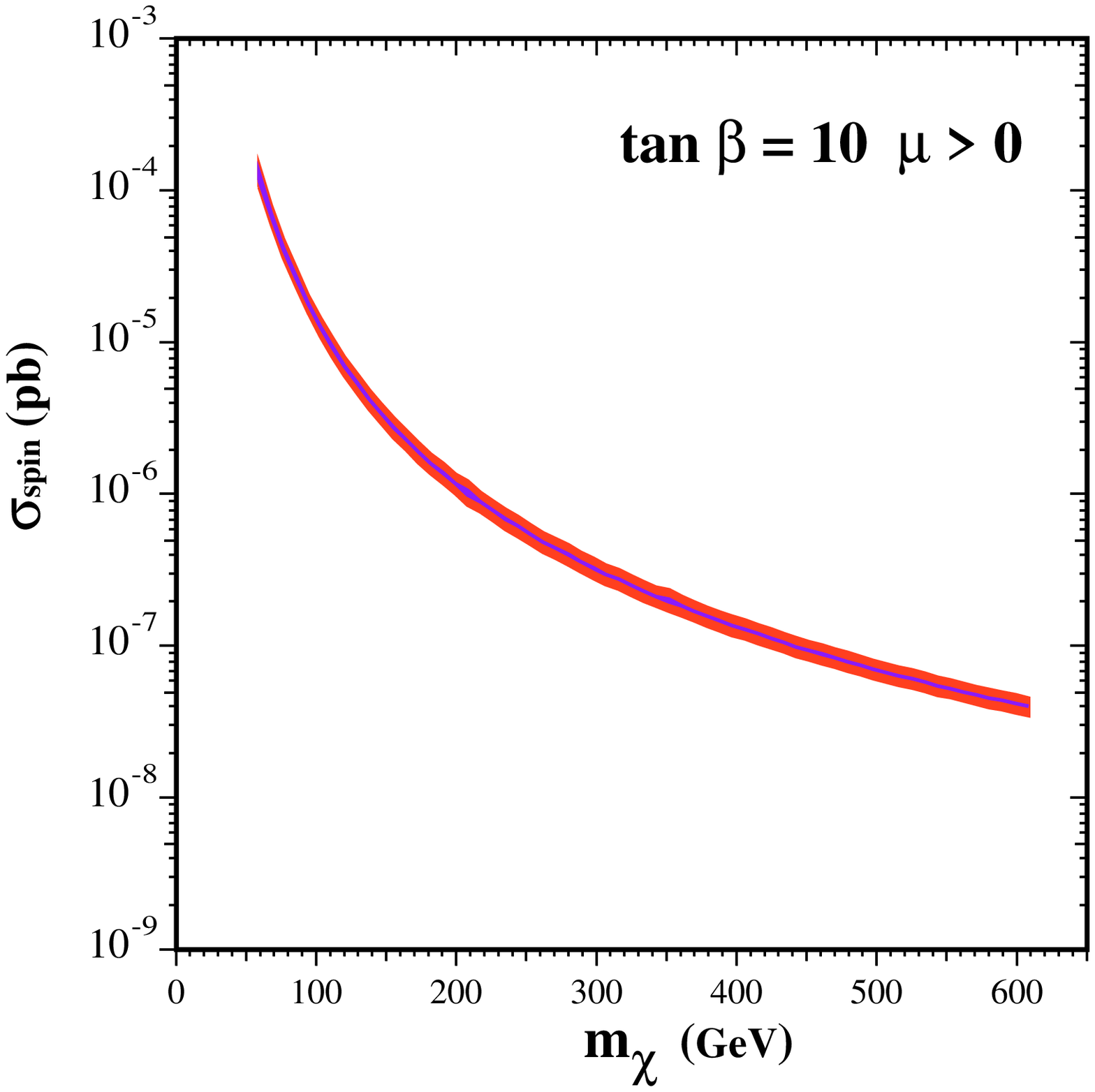,height=5.5in} \hfill
\end{minipage}
\vskip-0.7in
\caption{\label{fig:spin}
{\it The spin-dependent cross section for the elastic scattering of
neutralinos on protons as a function of the LSP mass.
The central curves are based on the inputs
(\protect\ref{spincontent}), and their thicknesses are related to
the spreads in the allowed values of $m_0$.  The shaded regions
correspond to the uncertainties in the hadronic inputs
(\protect\ref{spincontent}).}}
\end{figure}

In Figure 2, we show the corresponding result for the scalar cross
section, based on $\alpha_3$. As in Figure 1, the thickness of the central
curve reflects the range in $m_0$ sampled. The shaded region now
corresponds to the uncertainties in the inputs given in
(\ref{pinput}). The scalar cross section is, in general, more
sensitive to the sign of $\mu$ than is the spin-dependent 
cross section.
Notice that, in Figure 2c for $\tan \beta = 10$ and $\mu < 0$, there is
another cancellation. In this case, Higgs exchange is dominant in
$\alpha_3$. We first note that, for $\mu < 0$, both $Z_{\chi 3}$ and
$Z_{\chi 4}$ are negative, as is
the Higgs mixing angle $\alpha$. 
Inserting the definitions of $\delta_{1i(2i)}$, we see that there
is a potential cancellation of the Higgs contribution to $\alpha_3$ for
both up-type and down-type quarks. Whilst there is such a cancellation
for the down-type terms, which change from positive to negative as
one increases
$m_\chi$, such a cancellation does not occur for
the up-type terms, which remain negative in the region of parameters we
consider. The cancellation
that is apparent in the figure is due to the cancellation in
$\alpha_3$ between the up-type contribution (which is negative) and the
down-type contribution, which is initially positive but
decreasing, eventually becoming negative as we increase $m_{\chi}$.

\begin{figure}
\vspace*{-1.8in}
\hspace*{-.45in}
\begin{minipage}{8in}
\epsfig{file=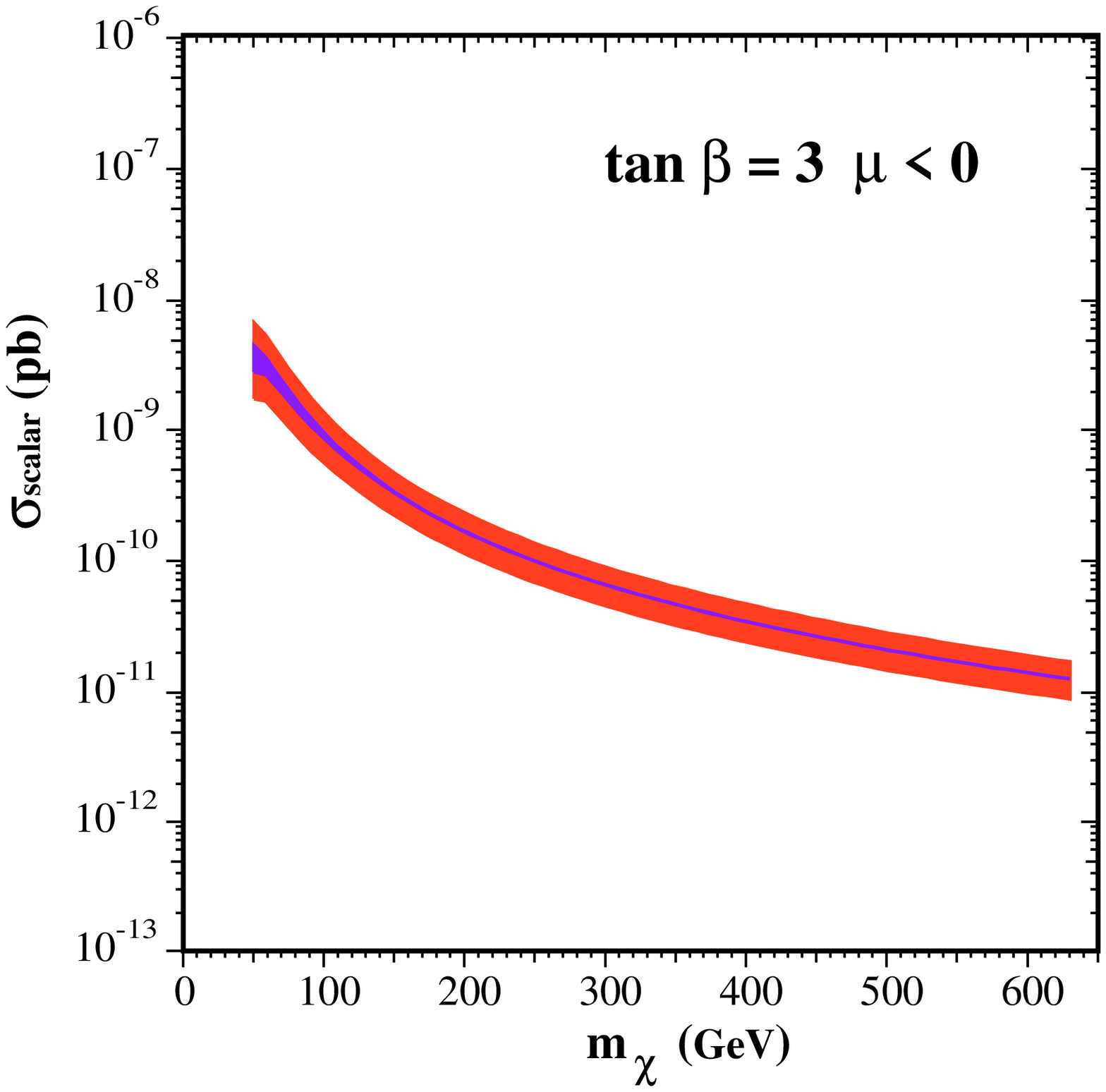,height=5.5in} 
\hspace*{-.40in}\epsfig{file=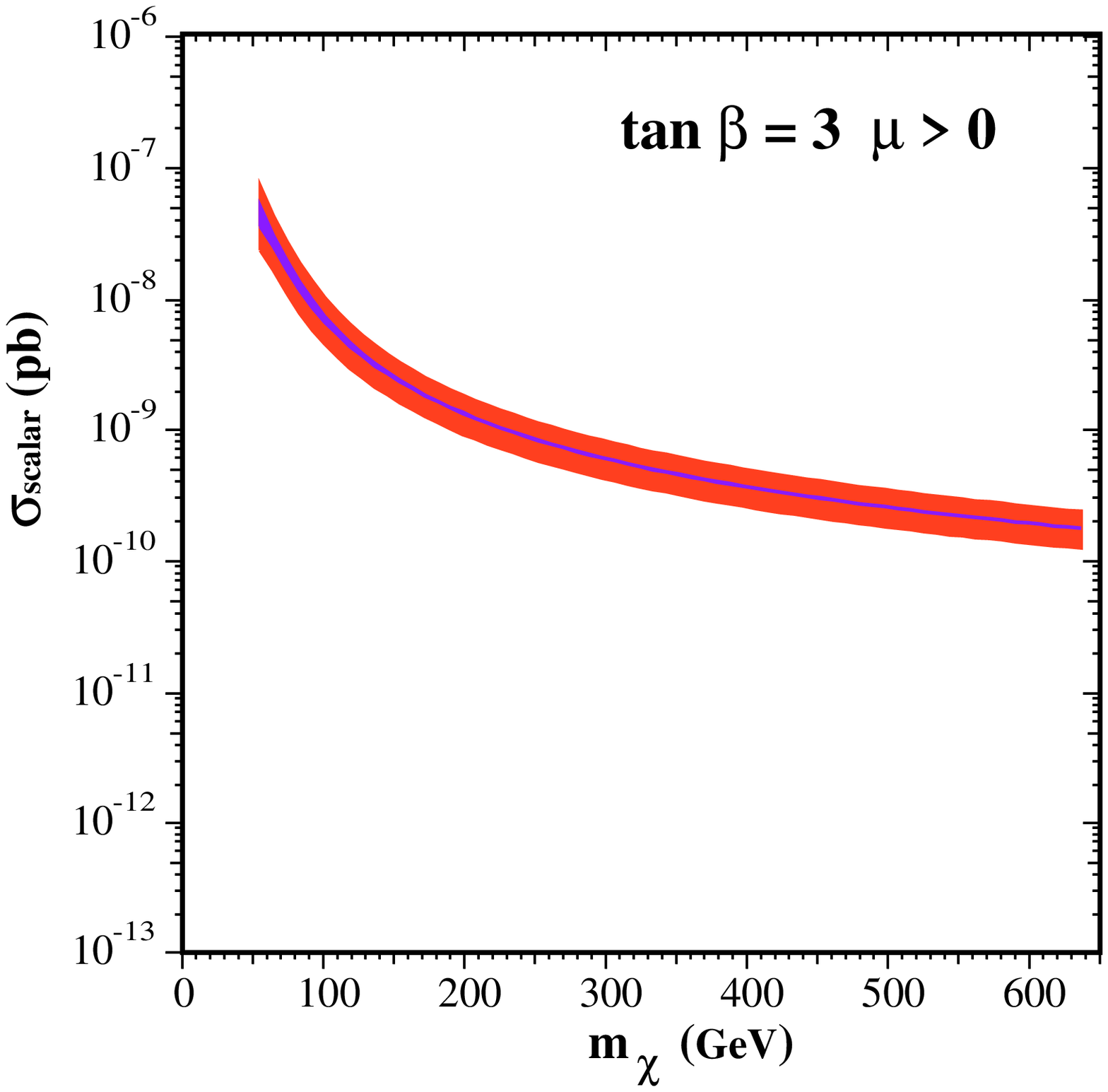,height=5.5in} \hfill
\end{minipage}
\begin{minipage}{8in}
\vspace*{-1.9in}
\hspace*{-.45in}
\epsfig{file=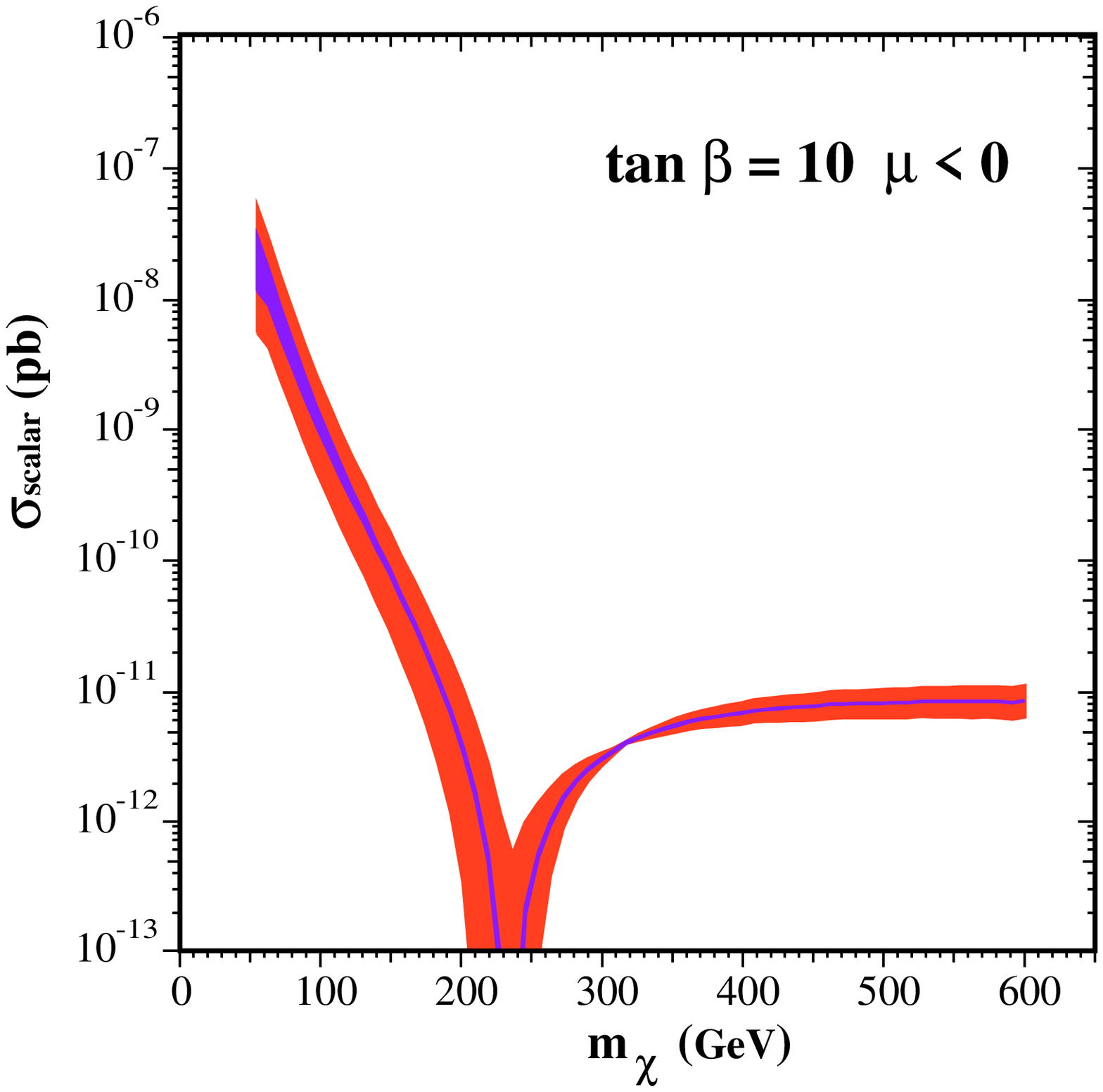,height=5.5in} 
\hspace*{-0.4in}
\epsfig{file=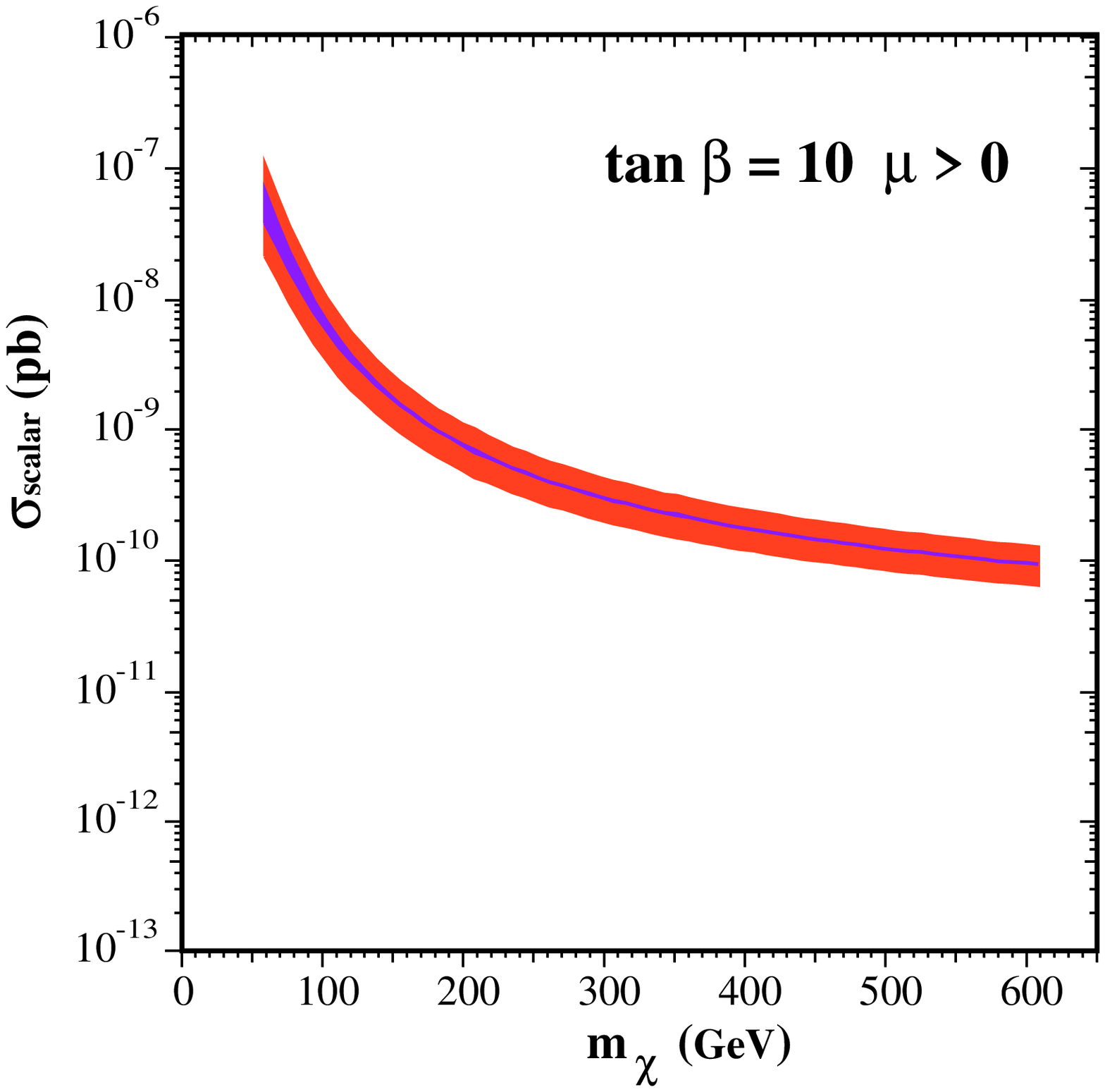,height=5.5in} \hfill
\end{minipage}
\vskip-0.7in
\caption{\label{fig:scalar}
{\it 
The spin-independent scalar cross section for the elastic scattering
of neutralinos on protons as a function of the LSP mass. The
central curves are based on the inputs (\protect\ref{pinput}),
their thicknesses are again related to the spread in the
allowed values of $m_0$, and the shaded regions now correspond to the
uncertainties in the hadronic inputs (\protect\ref{pinput}).}}
\end{figure}   

In Figure 3, we show the effects of varying some of the input assumptions
made earlier.  For example, when the assumed uncertainty in $y$ is taken
to be $\pm$0.2, we get a thicker shaded region, as shown for
$\tan \beta = 3, \mu < 0$ in Figure 3a. In Figure 3b, we give one example 
of the cross section for the elastic scattering of neutralinos on
neutrons. This particular case was chosen because it displays the largest
difference between the neutron and proton cross sections among those 
tested. As one can
see, our results for the LSP scattering on neutrons and protons are
almost identical.  Similarly, the effects of changing $A_0$ are also
relatively small, as illustrated by two cases with $A_0=0$ in
Figures 3c and 3d. In the latter example, there is almost a
factor of 2 difference at higher values of $m_\chi$, 
which is due to yet another cancellation, this time between
the squark-exchange and $Z$-exchange terms in $\alpha_{2u}$. 

\begin{figure}
\vspace*{-1.8in}
\hspace*{-.45in}
\begin{minipage}{8in}
\epsfig{file=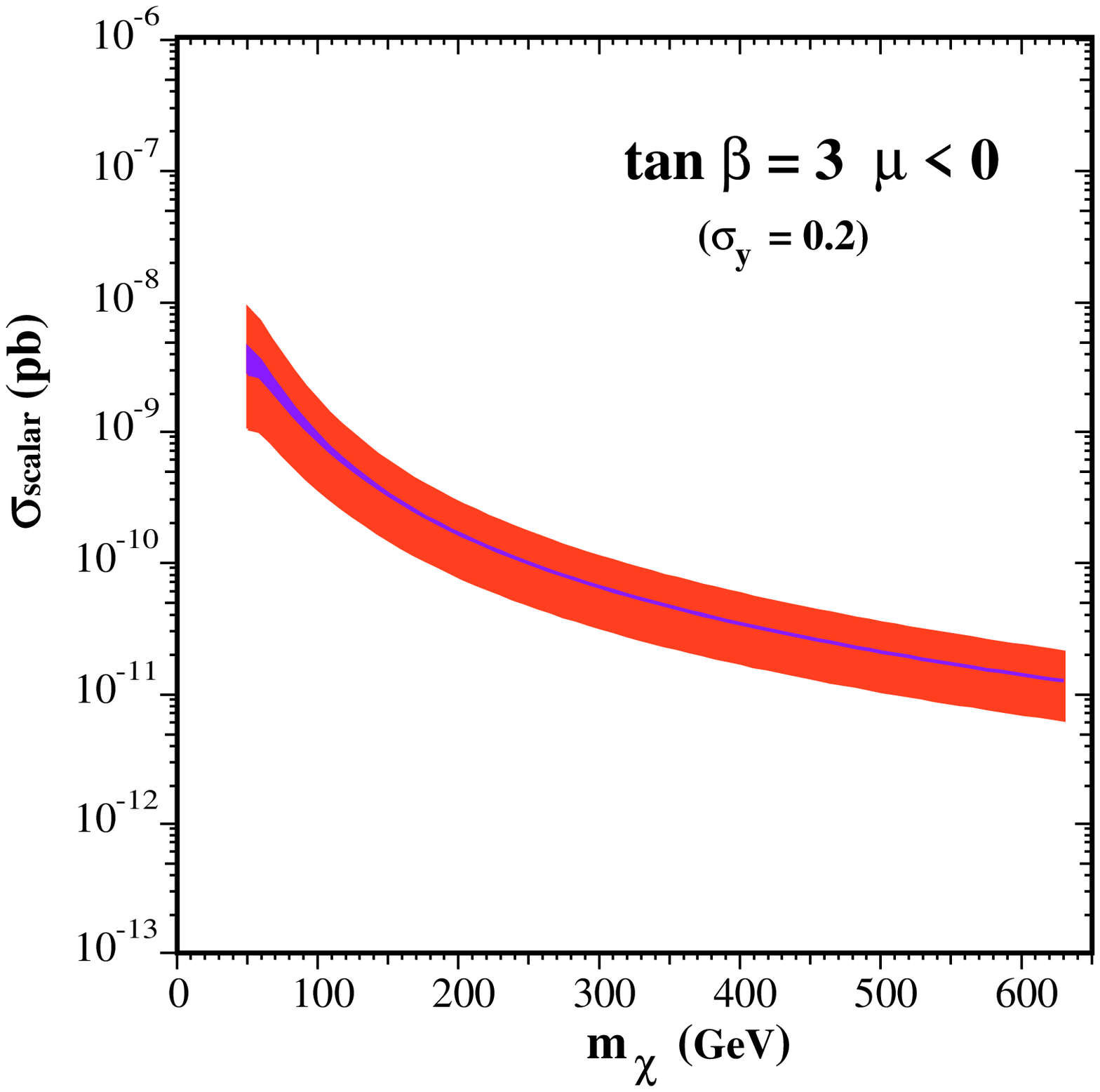,height=5.5in} 
\hspace*{-.40in}\epsfig{file=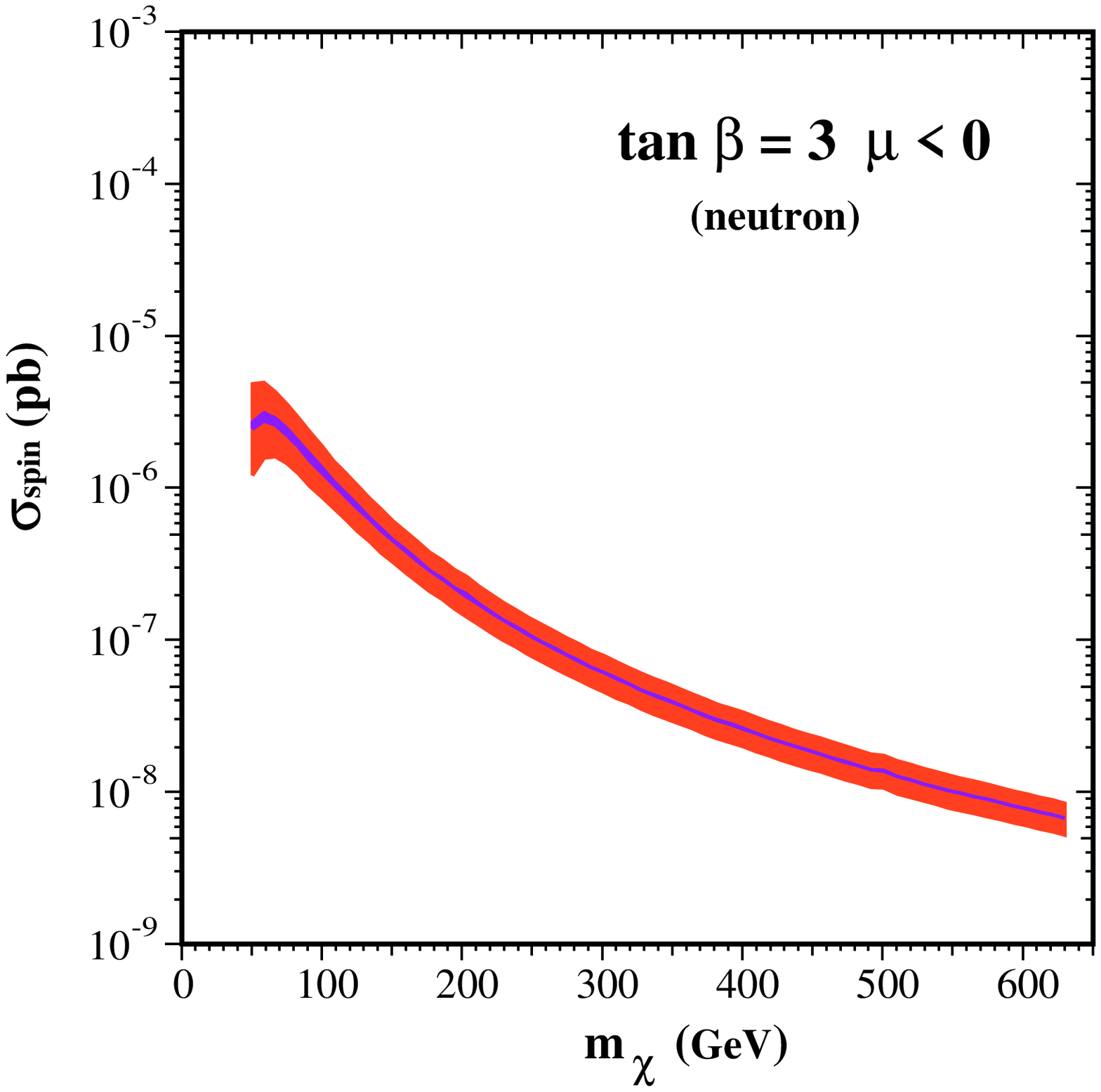,height=5.5in} \hfill
\end{minipage}
\begin{minipage}{8in}
\vspace*{-1.9in}
\hspace*{-.45in}
\epsfig{file=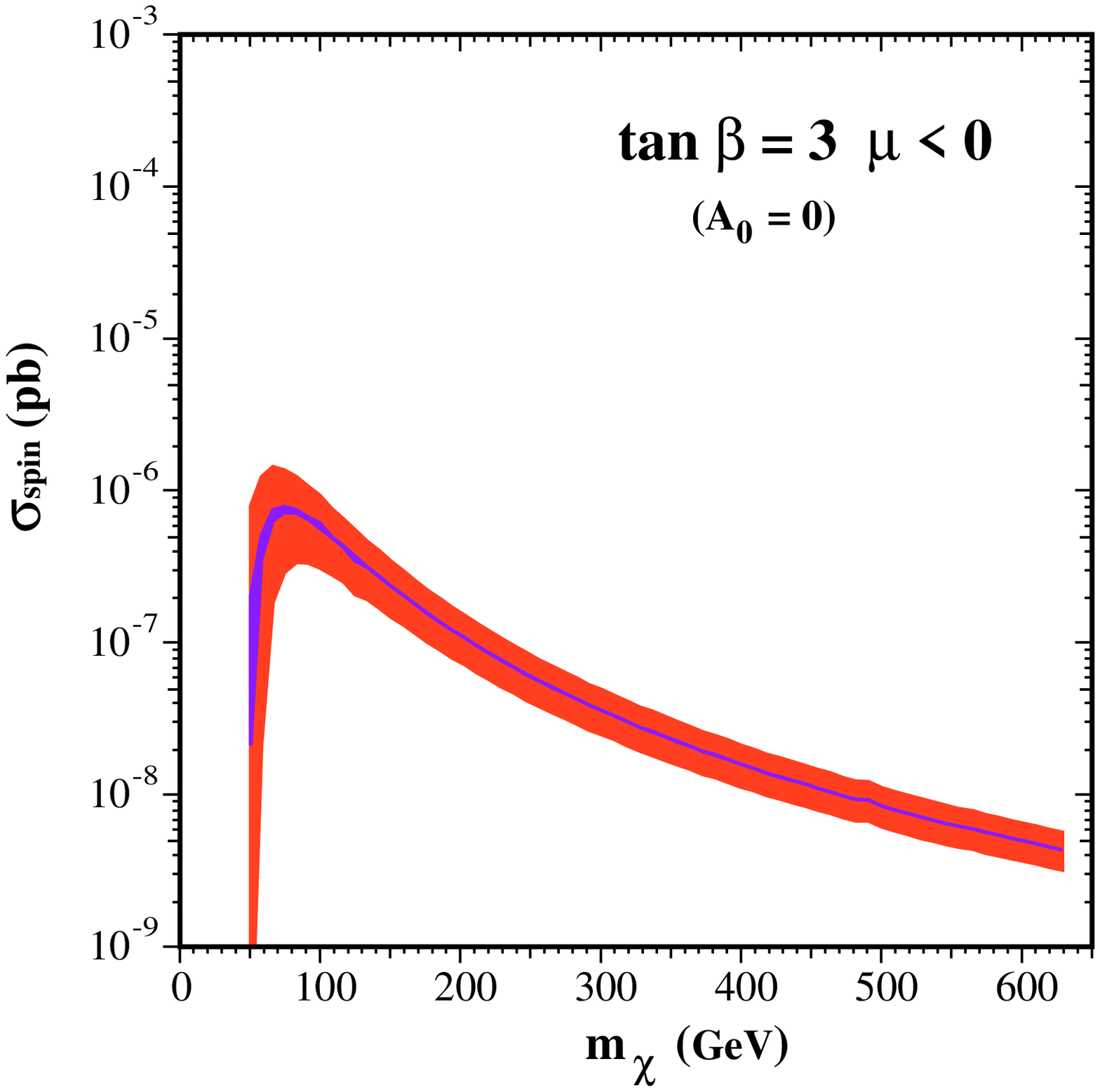,height=5.5in} 
\hspace*{-0.4in}
\epsfig{file=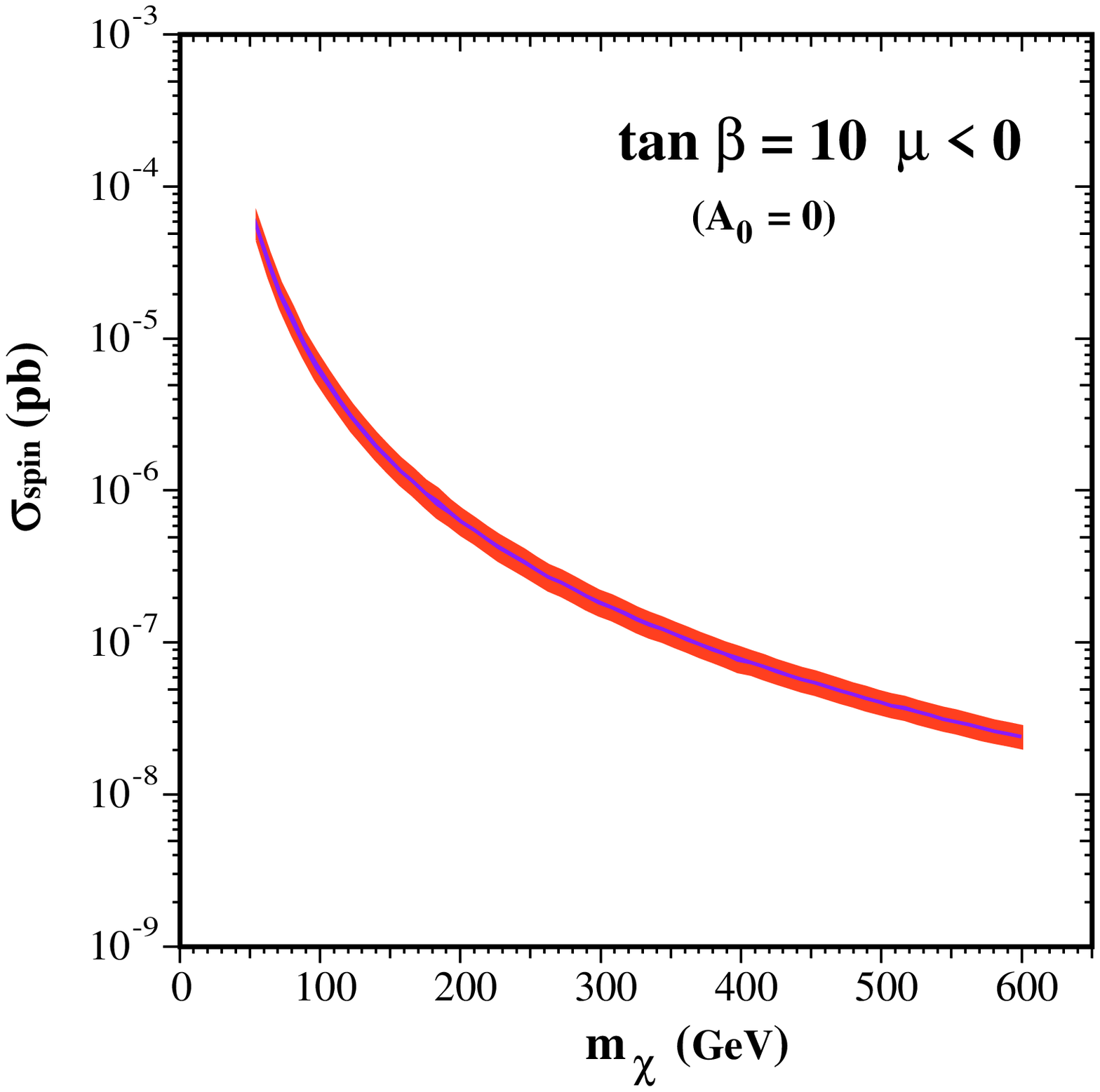,height=5.5in} \hfill
\end{minipage}
\vskip-0.7in
\caption{\label{fig:var}
{\it As in Figures 1 and 2, but now illustrating a) the effect  of
enlarging the
uncertainty in $y$ to $\pm 0.2$, b) the spin-dependent cross section for
the elastic
scattering of neutralinos on neutrons,  c) and d) the effects of setting
$A_0 = 0$ for $\mu < 0$ and two choices of $\tan \beta$.}}
\end{figure}

Finally, we show in Figure 4 compilations of our  results for the
spin-dependent and -independent
cross sections, compared with current and projected experimental
limits obtained from~\cite{uppersigma}. The shaded
region in panel (a) is the union of the shaded regions in Figure 1, and
the shaded region in panel (b) is the union of the shaded regions in
Figure 2. 

\begin{figure}
\vspace*{-1.8in}
\hspace*{-.45in}
\begin{minipage}{8in}
\epsfig{file=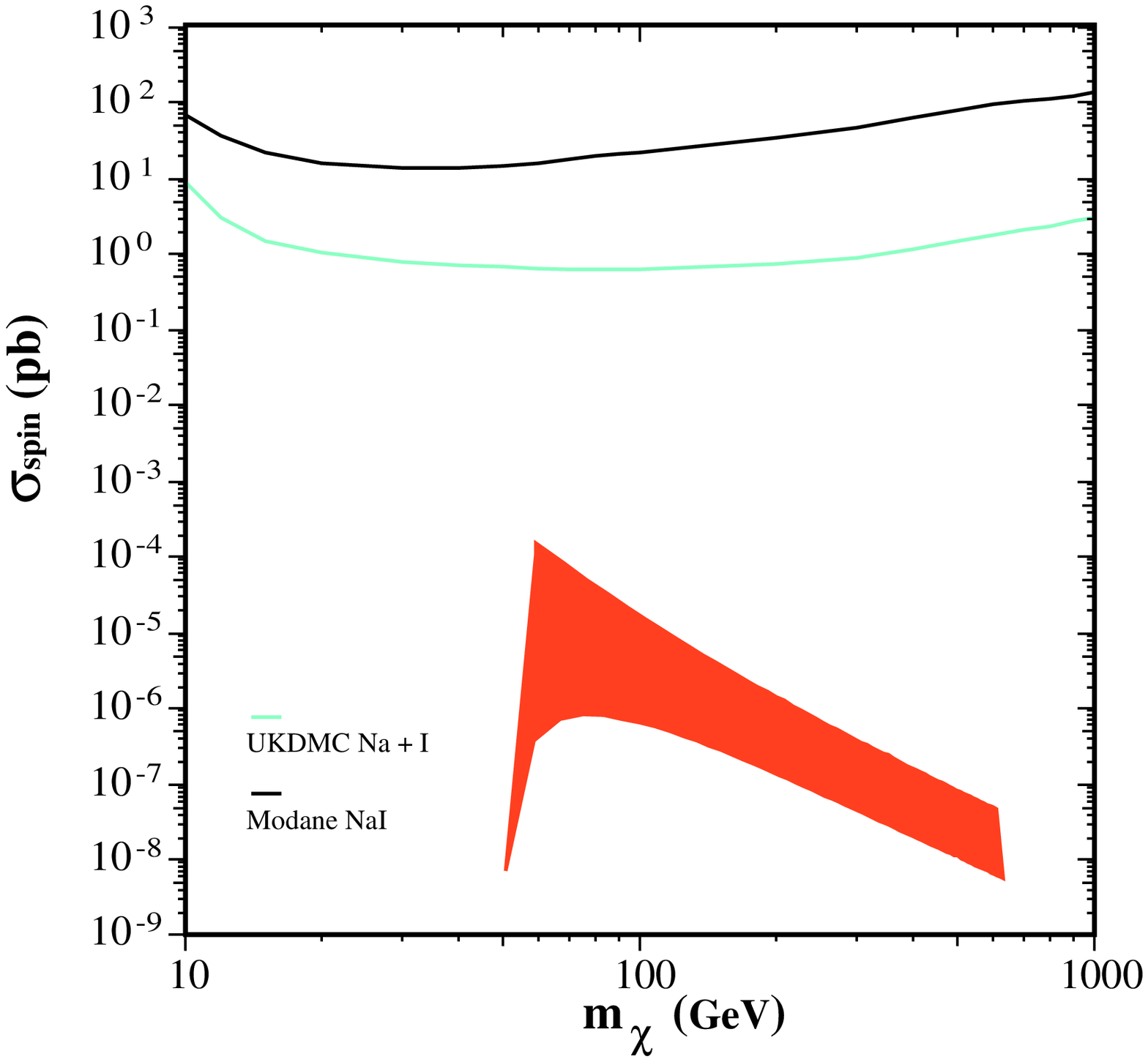,height=5.5in} 
\hspace*{-.40in}\epsfig{file=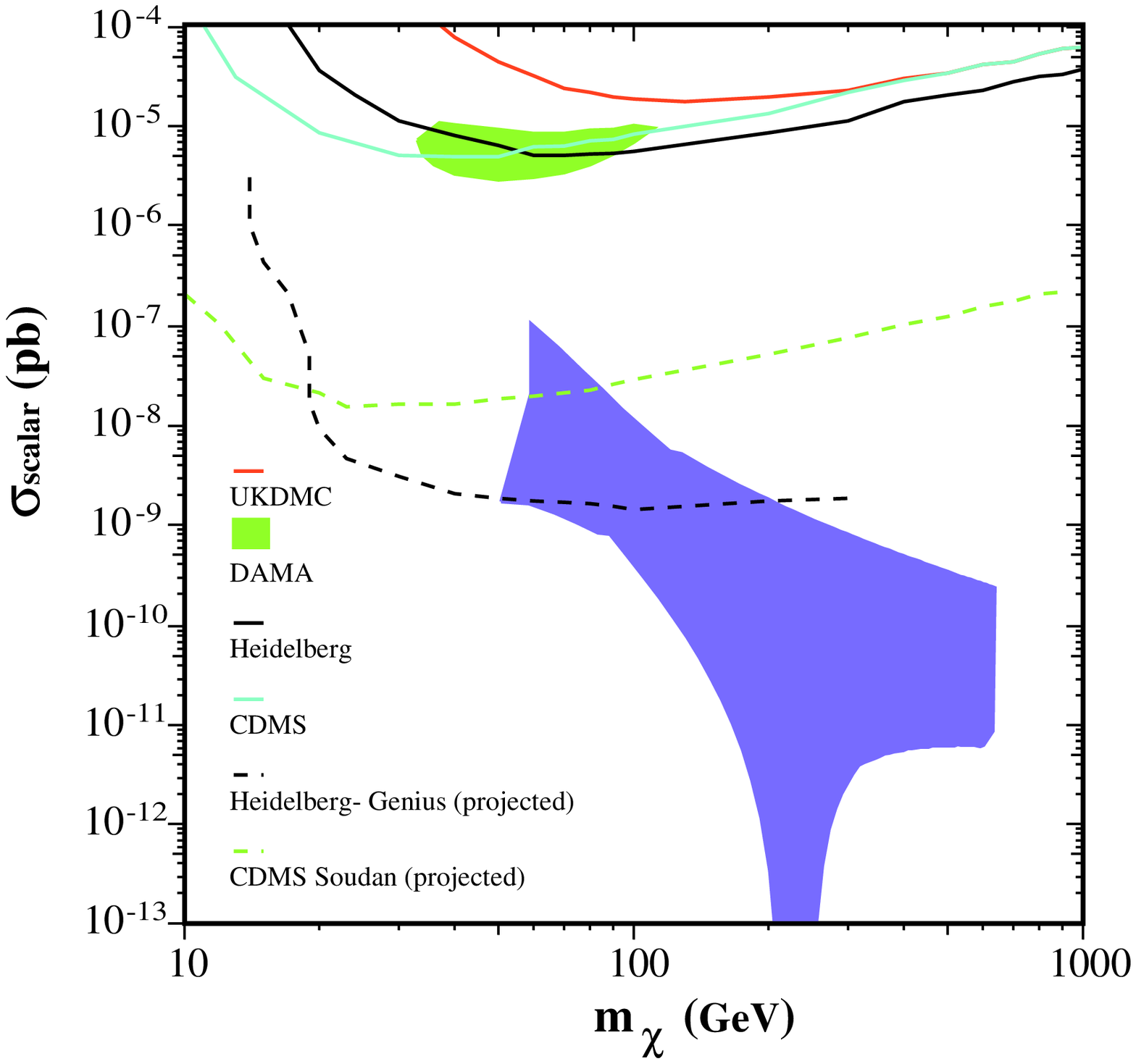,height=5.5in} \hfill
\end{minipage}
\vskip-0.7in
\caption{\label{fig:sm}
{\it Summary plot compiling our predictions in the constrained MSSM for 
(a) the spin-dependent and (b) the 
spin-independent elastic LSP scattering cross sections on protons,
compared in each case with current and projected limits on the
scattering cross sections, as obtained from~\cite{uppersigma}.}}
\end{figure}

\section{Discussion}

As seen in Figure~4, the present experimental upper
limit~\cite{uppersigma} on the spin-independent part of the 
elastic scattering of the LSP on a nucleon is around $10^{-5}$~pb
for $50~{\rm GeV} \la m_\chi \la 100$~GeV. 
On the other hand, the maximum scalar cross section we find
is around $10^{-8}$~pb, which is attained for $m_\chi \sim 50$~GeV.
This means that present experiments searching directly for supersymmetric
dark matter are far from constraining the parameter space of our baseline
theoretical framework, in which LEP constraints are applied to 
MSSM models with universal supergravity-inspired 
soft supersymmetry-breaking parameters $m_{1/2}, m_0$.

The literature contains predictions for the elastic LSP-nucleus
scattering rates that vary considerably, with some estimates lying
considerably higher than ours~\cite{GJK,bigguys}. There are various ways
in which
such differences might arise, of which we mention a few here. We have
imposed the requirement that the LSP relic density lie in the favoured
range $0.1 \le \Omega_\chi h^2 \le 0.3$, whereas other calculations
often include models with lower relic densities. Such models would
normally have larger $\chi \chi$ annihilation cross sections, and
correspondingly larger elastic scattering cross sections. Hence the
predicted scattering rates would be larger, if the conventional halo
density $\rho \sim 0.3$~GeV/cm$^3$ is assumed for the
LSP~\cite{uppersigma}. However, we believe this assumption is
unreasonable: if not all the total cold dark matter density $\Omega_{CDM}$ is
composed of LSPs, the density of LSPs in the halo should be reduced by the
corresponding factor $\Omega_\chi / \Omega_{CDM}$.

Other possible differences may arise in the treatment of the LEP constraints:
we find it to be almost excluded that the LSP be
Higgsino-like~\cite{EFGOS}, even if
the assumptions of universal soft supersymmetry breaking are relaxed, and
Higgsino dark matter is certainly excluded if universality is assumed, as is
the case here. In addition to the LEP constraints, this is because
the value of $\mu$ is predicted as a function of $m_{1/2}$ and $m_0$,
placing the LSP firmly in the Bino-like region. The same considerations
exclude an LSP with mixed Higgsino/gaugino content.

The prediction of $\mu$ may be circumvented by postulating non-universality
for the soft scalar supersymmetry-breaking parameters in the Higgs sector,
which might have appeared to resurrect the possibility of a Higgsino-like
LSP~\cite{Bottino}. However, such a possibility goes beyond the
universality framework adopted here, and, moreover, the LEP constraints
now appear to exclude this possibility~\cite{EFGOS}, as mentioned above.

There are no differences between the effective Lagrangians we and
others~\cite{EF,FFO1,FFO2,CIN}
use to describe the four-fermion $\chi - q$ interaction that
determines the elastic $\chi - p,n$ scattering cross sections.
We have found differences of detail between our 
and other treatments of the
hadronic matrix elements of the scalar and axial-current  ${\bar q} q$
operators appearing in this Lagrangian~\cite{newBottino}, but this is not
responsible for any big difference between the results.

We should not want our experimental colleagues to be too downcast
by the long road they appear to have to cover in order to probe the
minimal universal MSSM framework utilized here. For example, there
are surely some supersymmetric models that predict larger scattering rates.
However, we think it best to have in mind a plausible and realistic
target sensitivity, which is what our universal framework and
implementation of the LEP and cosmological constraints provide.
Our results also have the merit of being relatively specific: as seen in
Figure 4,
the elastic scattering cross sections we predict for any given value
of the LSP mass $m_\chi$ lie in a comparatively narrow band. As discussed
earlier, this is essentially because the LSP is always mainly
Bino-like in our framework, so its couplings do not depend
greatly on other MSSM parameters such as $m_0$. The principal causes
of broadening are the uncertainties in the hadronic inputs and the
possibilities of cancellations that may reduce the cross sections for
some specific values of the constrained MSSM parameters.

This tight correlation we find between the LSP mass and 
its elastic scattering rate means
that future experiments~\cite{Genius} should be able to phrase their
sensitivities
directly in terms of the LSP mass in the universal supergravity-inspired
version of the MSSM. For example, our results suggest that the
proposed Genius experiment~\cite{Genius} would be sensitive to
$m_\chi \la 100$~GeV for almost all MSSM parameter choices in Figure 5b.
More optimistically, if/when a signal is observed, its
plausibility would be enhanced if its recoil spectrum was correlated with
the rate in the manner suggested by Figure 5. Thus our analysis
provides experiments with an additional tool that may assist in the
extraction of a signal that might be significantly smaller than they could
have hoped. In any case, the importance of the search for supersymmetric
matter remains unchanged, and there are still several years before the LHC
comes into operation, so these experiments still have both motivation
and opportunity.

\vskip 0.5in
\vbox{
\noindent{ {\bf Acknowledgments} } \\
\noindent 
We thank Toby Falk and Gerardo Ganis for many related discussions. 
The work of K.A.O. was supported in part by DOE grant
DE--FG02--94ER--40823.}

\end{document}